\documentclass[twoside]{article}
\usepackage{fleqn,espcrc2}
\usepackage{amsbsy,epsf,amssymb}
\newcommand{\bq}{\boldsymbol {q}}
\newcommand{\bk}{\boldsymbol {k}}

\title{Self-consistent description of simplified pseudospin-electron model}
\author{{I.V.~Stasyuk, A.M.~Shvaika and K.V.~Tabunshchyk}
\address{Institute for Condensed Matter Physics NASU,
            1 Svientsitskii Str., UA--79011 Lviv, Ukraine}}
\begin{document}
\begin{abstract}
 A method of the self-consistent calculation of the
thermodynamical and correlation functions is presented.
 This approach is based on the GRPA (generalized random phase
approximation) scheme with the inclusion of the mean field
corrections.
 The numerical research shows that interaction between the electron and
pseudospin subsystems leads to the possibility of the dipole
(pseudospin) and charge-density instabilities as well as phase
separation into the uniform and/or chess-board phases.
\end{abstract}
\maketitle

 The aim of this article is to calculate the correlation
functions, pseudospin and particle number mean values as well as
the grand canonical potential for the simplified ($U=0$)
pseudospin-electron model (PEM) \cite{Shvaika} within the one
self-consistent approach.

 PEM includes the local interaction of the conducting electrons with
some two level subsystem described by pseudospins $S^z{=}\pm 1/2$
(e.g anharmonic vibrations of the apex oxygen ions in YBaCuO-type
crystals) \cite{Muller}.
 On the other hand, PEM can be transform into the binary alloy type
model as well as into the Falikov-Kimball model (FKM) and they can
be also studied analytically within the framework of the developed
scheme.

  The model Hamiltonian is the following:
\begin{eqnarray}
 \label{Hamiltonian1}
\hspace*{-.7cm}
 &&H{=}\sum_iH_{i\,0}+\sum_{ij\sigma} t_{ij}c_{i\sigma}^{+} c_{j\sigma},\\
\hspace*{-.7cm}
 &&H_{i\,0}{=}
   {-} \mu\sum_{\sigma}n_{i\sigma}
   {+} g\sum_{\sigma} n_{i\sigma} S_{i}^{z} {-} hS_{i}^{z} ,\nonumber
\end{eqnarray}
where an interaction with pseudospins ($g$-term) placed in some
longitudinal field $h$ (chemical potential for ions in FKM) are
included in the single-site part; $\mu$ is the chemical potential.
 The second term in the Hamiltonian describes an electron hopping
from site to site.

 The calculations are performed in the strong coupling case ($g\gg t$) using
single-site states as the basic one.
 A formalism of the electron annihilation (creation) operators
$
 a_{i\sigma}=c_{i\sigma}P^+_i,
$
$
 \tilde{a}_{i\sigma}=c_{i\sigma}P^-_i
$
($P^{\pm}_i=\frac 12\pm S^z_i$) acting at a site with the certain
pseudospin orientation is introduced.

 Expansion of the calculated quantities in terms of the electron transfer
leads to an infinite series of terms containing the averages of
the $T$-products of the $a_{i\sigma}$, $\tilde{a}_{i\sigma}$
operators.
 The evaluation of such averages is made using the corresponding
Wick's theorem.
 The results are expressed in terms of the products
of the nonperturbed Green's functions and averages of the
projection operators $P^{\pm}_i$ which are calculated by means of the
semi-invariant expansion \cite{Tabunshchyk}.

 The calculation of the correlation functions is performed
within a self-consistent scheme in the framework of the
generalized random phase approximation \cite{Izyumov} with the
inclusion of the mean field type contributions coming from the
effective pseudospin interaction via conducting electrons, i.e.
all zero-order correlators (second order semi-invariants) as well
as pseudospin mean value are calculated by the mean-field
Hamiltonian:
\begin{eqnarray}
&&\hspace*{-.8cm} H_{\rm MF}=
   \sum\limits_iH_{i\,0}+\alpha_1P^+_i+\alpha_2P^-_i,\hspace*{-3cm}\\
&&\hspace*{-.8cm} \alpha_1P^+_i{+}\alpha_2P^-_i{=}
    \frac{2}{N\beta}\sum_{n,\bk}\frac{t^2_{\bk}}{g_n^{-1}{-}
t_{\bk}}\!\!\left[\frac{P^+_i}{{\rm i}\omega_n{-}\varepsilon_1}{+}
\frac{P^-_i}{{\rm i}\omega_n{-}\varepsilon_2}\right]\hspace*{-3cm}
 \nonumber\\
&&\hspace*{-.8cm} g_n = \frac{\langle P^+\rangle}
 {{\rm i}\omega_n-\varepsilon_1}+
 \frac{\langle P^-\rangle}{{\rm i}\omega_n-\varepsilon_2},
 \quad \varepsilon_{1,2}=-\mu\pm\frac g2.\hspace*{-3cm}
\end{eqnarray}

 Diagram equation on pseudospin correlator $\langle
S^zS^z\rangle_{\bq}$ is following:
\[
 \epsfysize 1.1cm\epsfbox{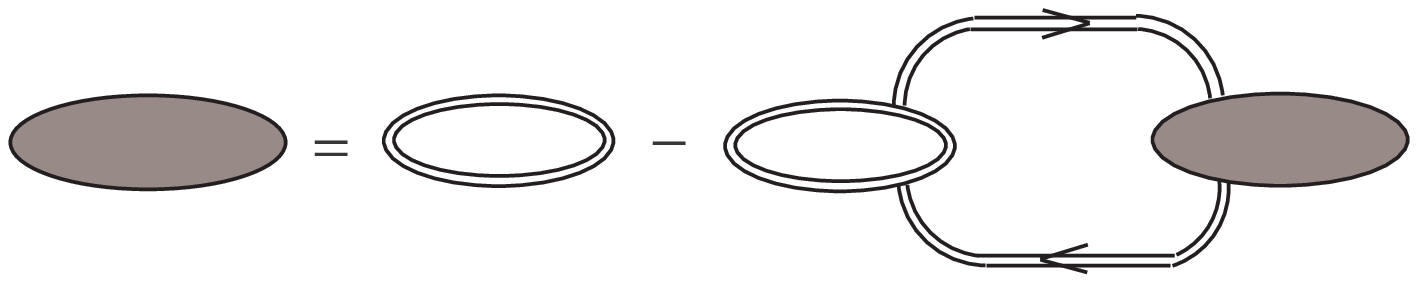}
\]
  This equation differs from the one for the Ising model
in MFA by the replacement of the exchange interaction by the
electron loop (which describes an interaction between pseudospins
via conducting electrons).
 In the analytical form its solution is equal
\begin{eqnarray}
 &&\chi^{S^zS^z}(\omega_m,\bq)=
\frac{\delta(\omega_m)\langle P^+\rangle\langle P^-\rangle}
     {T-\Theta(T,\bq)},\\
 && \Theta (T,\bq)=-\frac{2}{\beta}\sum_n\frac 1N\sum_{\bk}
 \Lambda^2_n\tilde{t}_n(\bk)\tilde{t}_n(\bk+\bq),\nonumber\\
 &&\Lambda_n{=}\frac{g\sqrt{\langle P^+\rangle\langle P^-\rangle}}
 {({\rm i}\omega {+}\mu)^2{-}g^2/4},
 \quad \tilde{t}_n(\bk){=}\frac{t_{\bk}}{(1{-}g_nt_{\bk})}.\nonumber
\end{eqnarray}

 Equation for pseudospin mean value in the uniform case
($\langle S^z_i\rangle=\langle S^z\rangle$) is following:
\[
\langle S^z\rangle{=} \frac 12
\tanh\left\{\frac{\beta}{2}(h{+}\alpha_2{-}\alpha_1){+}
\ln{\frac{1{+}{\rm e}^{-\beta\varepsilon_1}}
 {1{+}{\rm e}^{-\beta\varepsilon_2}}}\right\}
\]
 We also calculate particle number mean value and grand canonical
potential \cite{Tabunshchyk}.
 All quantities can be derived from the grand canonical
potential by
\[
 \frac{{\rm d}\Omega}{{\rm d}(-h)}
 =\langle S^z\rangle,\quad
 \frac{{\rm d}\langle S^z\rangle}{{\rm d}(\beta h)}
 =\langle S^zS^z\rangle_{{\bq}=0},
\]
which show the thermodynamical consistence of the proposed
approximations.

 The analysis of the pseudospin correlator temperature behaviour
shows that high temperature phase become unstable with respect to
fluctuations with wave-vector ${\bq}=(\pi,\pi)$ (for some model
parameters values) that indicates the possibility of the phase
transition into a modulated (chess-board) phase.
 On the other hand, the phase transition between the uniform phases
with different pseudospin mean values (bistability) is
possible~\cite{Tabunshchyk}.

 From the comparison of the grand canonical
potential $\Omega$ values for uniform and chess-board phases ($\mu
{=}{\rm const}$ regime), the ($\mu ,h$) phase diagram is obtained
(Fig.~1a).
\begin{figure}
 \raisebox{-0.15\textwidth}[2.3cm][2.2cm]
 {\epsfxsize 0.31\textwidth\epsfbox{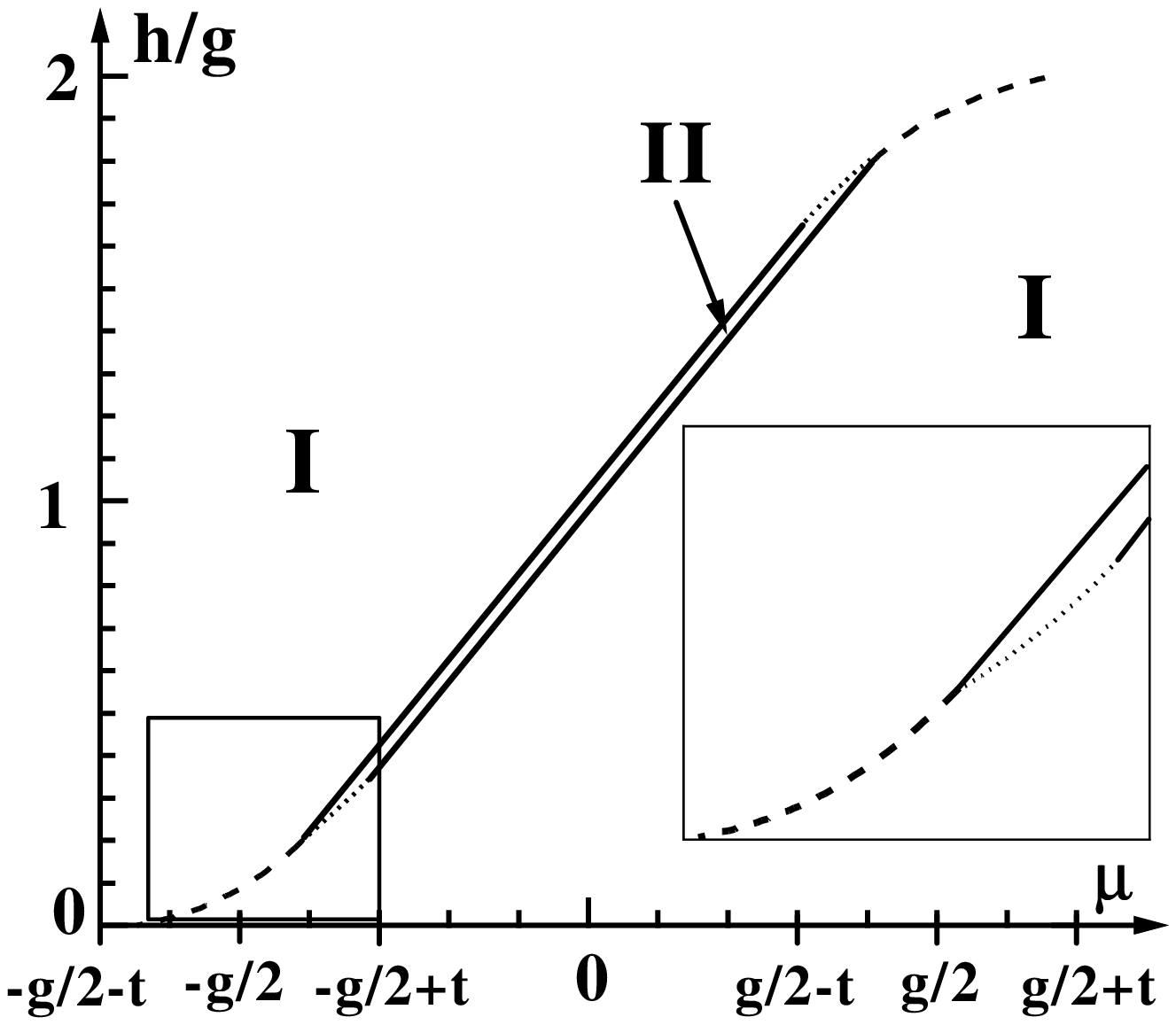}}\quad (a)\\
 $\;\;$\raisebox{-0.15\textwidth}[2.25cm][1.8cm]
 {\epsfxsize 0.29\textwidth\epsfbox{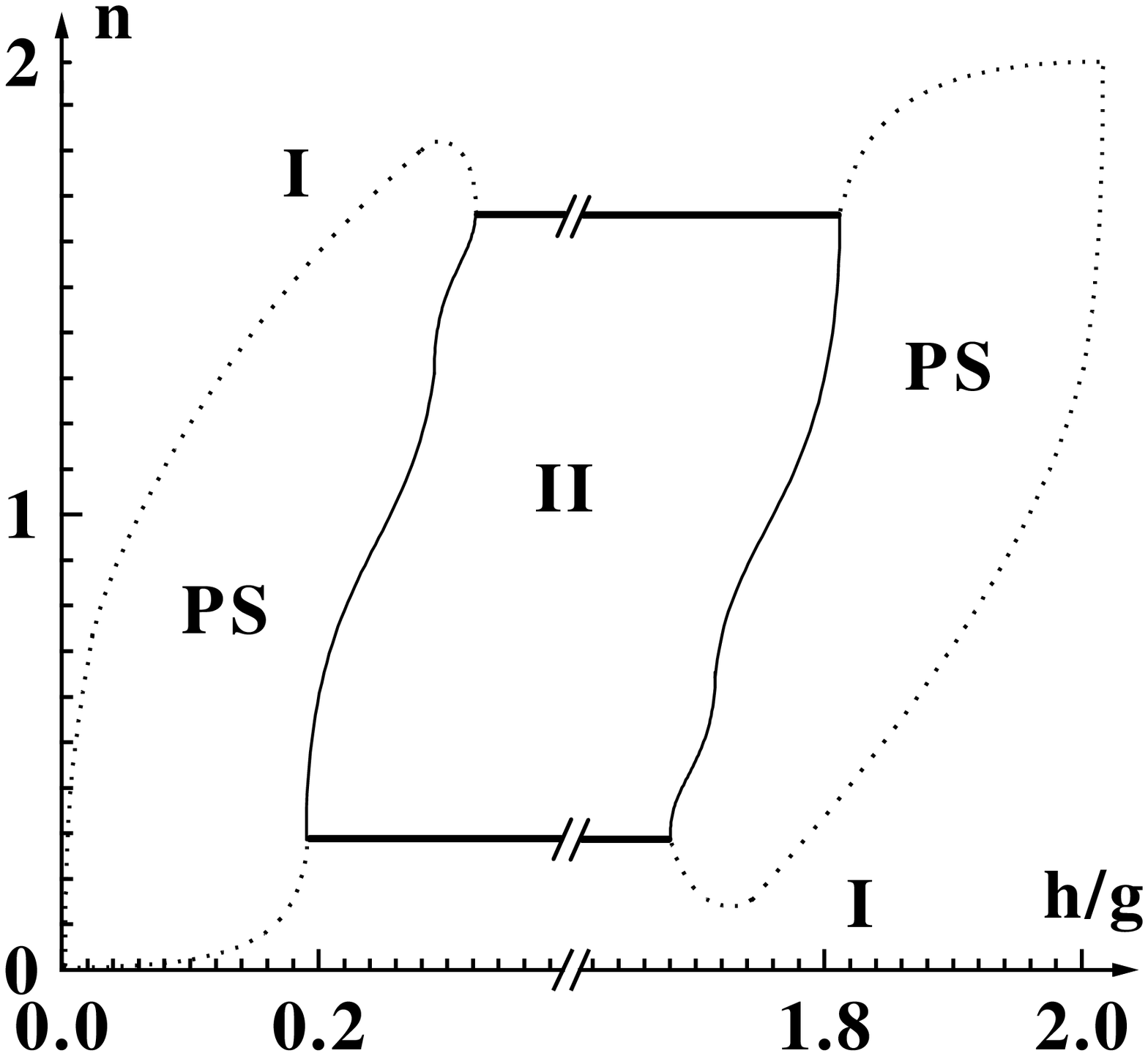}}\quad (b)
\caption{Phase diagrams: (a) $\mu-h$, (b) $n-h$.
 I -- uniform phase, II -- chess-board phase,
 PS -- phase separation area. ($T/g=0.005$, $t/g=0.2$)}
 \label{Fig1}
 $$\vspace*{-1.7cm}$$
\end{figure}

 Chess-board phase exists as intermediate one
between the uniform phases with different $\langle S^z\rangle$ and
$\langle n\rangle$ values. The transition between different
uniform phases (bistability) is of the first order (Fig.~1a,
dashed line), while the transition between the uniform and
modulated ones is of the first (dotted line) or second (solid
line) order.

 On the other hand, the minimum of the free
energy $F{=}\Omega {+} \mu N$ is the equilibrium condition in the
$n{=}{\rm const}$ regime.
 In this regime the first
order phase transition with a jump of the pseudospin mean value
accompanied by the change of electron concentration transforms
into a phase separation into the regions with different phases
(the uniform and chess-board ones) and with different electron
concentrations and pseudospin mean values~(Fig.~1b).

 On the basis of the presented scheme the
thermodynamics of phase transitions, possibility of the phase
separations as well as appearance of the chess-board phase have
been investigated.

%
%

\end{document}